\documentclass[aps,pre,superscriptaddress,twocolumn]{revtex4}

\usepackage{graphicx,amssymb,amsfonts,amsmath,chemarr,color,commath,hyperref}
\usepackage{xurl}
\usepackage[version=4]{mhchem}
\usepackage{float}

\usepackage[dvipsnames]{xcolor}

\begin{document}
\title{Deduction of signaling {mechanisms} from cellular responses to multiple cues}

\author{Soutick Saha}
\affiliation{Department of Physics and Astronomy, Purdue University, West Lafayette, IN 47907, USA}

\author{Hye-ran Moon}
\affiliation{School of Mechanical Engineering, Purdue University, West Lafayette IN 47907, USA}

\author{Bumsoo Han}
\affiliation{School of Mechanical Engineering, Purdue University, West Lafayette IN 47907, USA}
\affiliation{Purdue Center for Cancer Research, Purdue University, West Lafayette, IN 47907, USA}

\author{Andrew Mugler}
\email{andrew.mugler@pitt.edu}
\affiliation{Department of Physics and Astronomy, Purdue University, West Lafayette, IN 47907, USA}
\affiliation{Purdue Center for Cancer Research, Purdue University, West Lafayette, IN 47907, USA}
\affiliation{Department of Physics and Astronomy, University of Pittsburgh, Pittsburgh, PA, 15260, USA}

\begin{abstract}
Cell signaling networks are complex and often incompletely characterized, making it difficult to obtain a comprehensive picture of the mechanisms they encode. Mathematical modeling of these networks provides important clues, but the models themselves are often complex, and it is not always clear how to extract falsifiable predictions. Here we take an inverse approach, using experimental data at the cell level to {deduce} the minimal signaling network. We focus on cells' response to multiple cues, specifically on the surprising case in which the response is antagonistic: the response to multiple cues is weaker than the response to the individual cues. We systematically build candidate signaling networks one node at a time, using the ubiquitous ingredients of (i) up- or down-regulation, (ii) molecular conversion, or (iii) reversible binding. In each case, our method reveals a minimal, interpretable signaling mechanism that explains the antagonistic response. Our work provides a systematic way to {deduce} molecular mechanisms from cell-level data.
\end{abstract}

\maketitle

Cell signaling networks are dauntingly complex. Increasingly precise biochemical experiments have characterized the structure of signaling networks in exquisite detail \cite{michal2013biochemical, icard2012global}. The availability of such a large amount of quantitative data suggests that it should be possible to understand the function of these networks and predict responses at the cellular level. However, the sheer complexity of cell signaling networks makes intuitive understanding and unambiguous prediction elusive.

One approach to the problem of signaling network complexity is to appeal to mathematical modeling. Deterministic dynamical models turn experimental information about network topology and kinetic parameters into predictive information about network function and cell response \cite{le2015quantitative, letunic2008ipath, medema2012computational}. However, models that respect the known degree of experimental detail are as complex as the networks themselves, by definition. Complex models are difficult to interpret intuitively, leaving the basic functional mechanisms underlying cell response unknown. Moreover, kinetic parameters are rarely known for all interactions, leaving a vast parameter space to explore theoretically \cite{zamora2011efficient}, and therefore many possible---often conflicting---predictions for cell response. Finally, experimental characterization of signaling networks is inevitably incomplete, and it is unclear whether a detailed model informed by current data will make robust predictions when updated with new data.

\begin{figure}
\centering
\includegraphics[width=\linewidth]{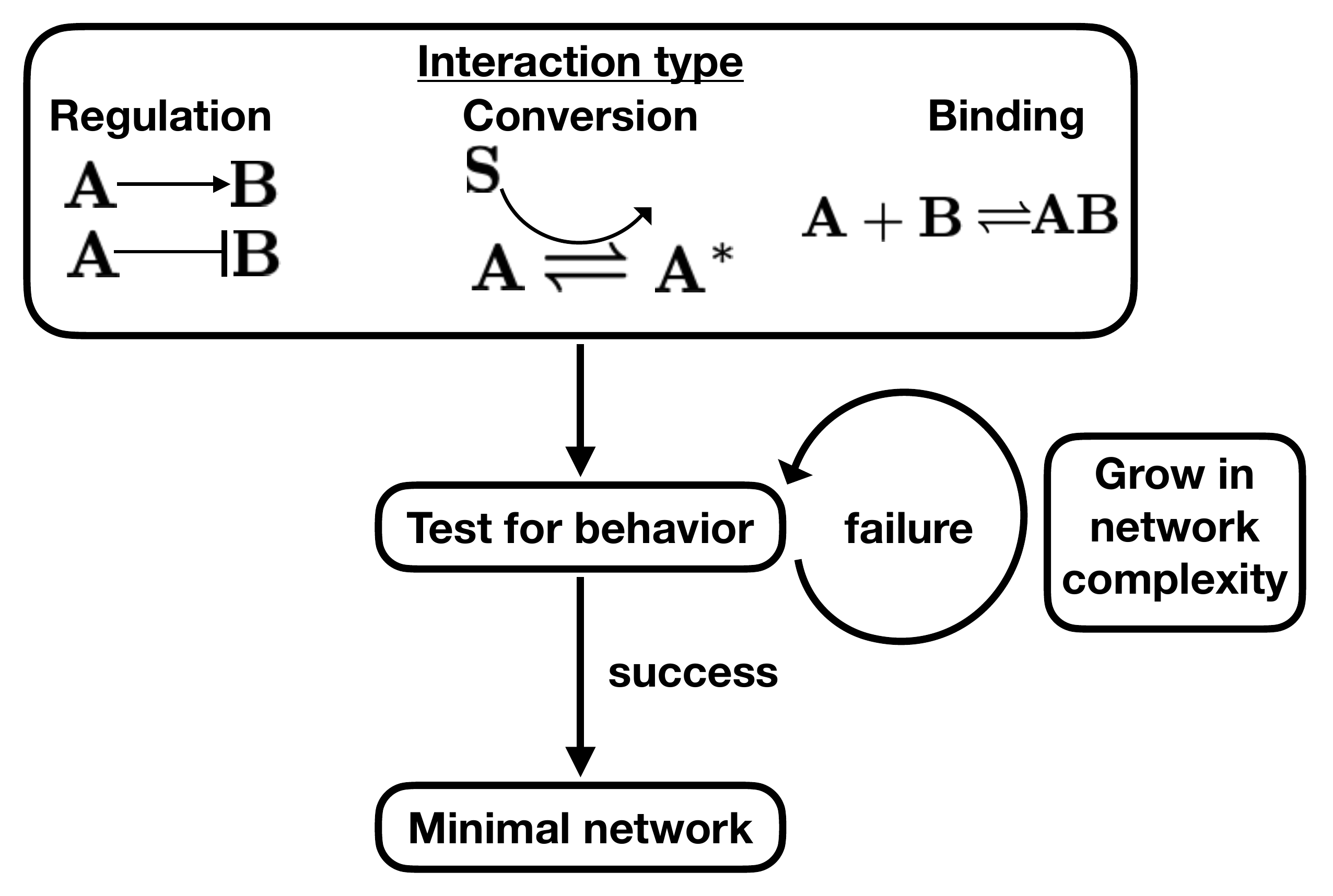}
\caption{\textbf{Workflow of method.} For each mechanism of (i) regulation, (ii) molecular conversion, or (iii) reversible binding, we start with the minimal possible network in terms of both nodes and edges and evaluate whether the network can explain the cell behavior. If the minimal network cannot explain the observed behavior we increase the complexity of the network by increasing the number of nodes or edges, one at a time, and we evaluate the new network. We repeat until the minimal successful network is identified.}
\label{method}
\end{figure}

Here we introduce an inverse approach to understanding complex signaling networks. Instead of modeling all known features of a signaling network to predict a cell response, we use observed cell responses to {deduce} a minimally sufficient signaling network \cite{franccois2007deriving}. Our approach is systematic: given a particular class of biochemical interactions, we build candidate networks one piece at a time, continually evaluating whether the observed set of cell responses is compatible with the current candidate network (Fig.\ \ref{method}). Networks therefore grow in complexity in a principled way, and we are left with the minimally complex network {or a set of equally complex minimal networks} of a given class that can explain the observed behavior. These minimal networks are not meant to have a one-to-one correspondence with the known experimental features of the cell signaling network. Rather, we conjecture that they reveal the coarse-grained structure of the signaling network: the gross topological features that are needed in order to implement the observed cell responses. Importantly, as we will show, working at this level provides intuition about why the topology encodes the function, and provides predictions that are robust to parameter changes, due to the small number of degrees of freedom.

{Iterative or exhaustive approaches have been developed in the past to uncover the common features in networks that perform temporal functions, such as signal adaptation \cite{ma2009defining} or robust oscillations \cite{wagner2005circuit}; spatial functions, such as embryonic patterning \cite{ma2006robustness, cotterell2010atlas, schaerli2018synthetic}; or information processing functions, such as noise reduction \cite{hornung2008noise}. Related approaches have derived simple networks via in-silico evolution in the service of particular fitness goals \cite{franccois2007deriving, franccois2014evolving}. Many of these works have focused on the performance of a specific function in response to a single input, whereas here we are interested in deducing minimal networks that underlie a system's ability to multiplex, i.e., to respond to multiple inputs either individually or together.}

\begin{figure}
\centering
\includegraphics[width=0.95\linewidth]{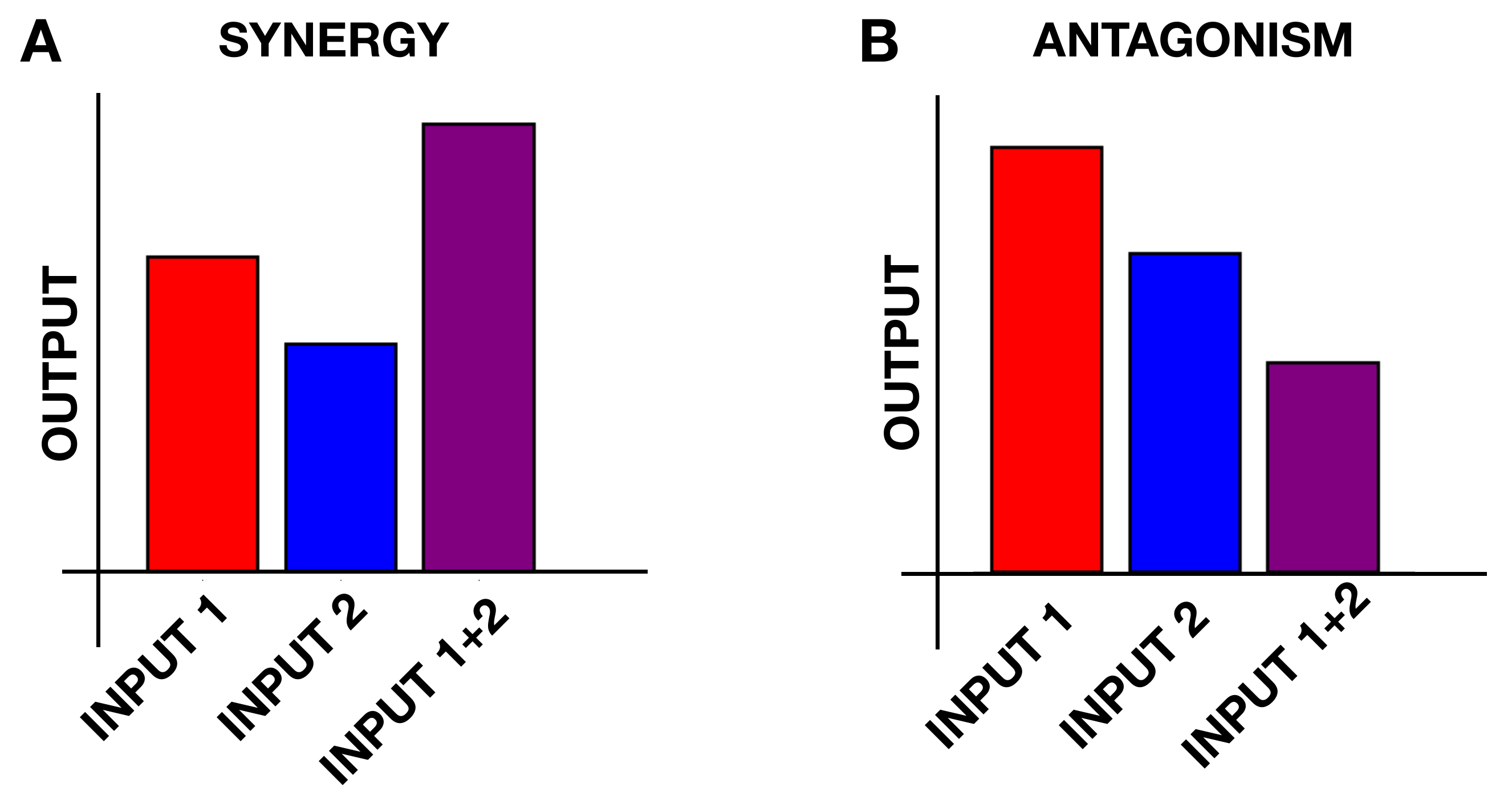}
\caption{\textbf{Synergy vs.\ antagonism.} (A) A synergistic response to two inputs is larger than the response to either input alone. (B) An antagonistic response to two inputs is smaller than the response to either input alone. Here we focus on antagonism.}
\label{antagonism}
\end{figure}

We focus on multi-stimulus behaviors where the cell response is antagonistic. This class of experiment involves stimulating a cell with two inputs, first one-at-a-time, then both simultaneously, and measuring its output \cite{badache2001interleukin, zhou2007synergistic, mosadegh2008epidermal, uttamsingh2008synergistic, buonato2015egf, schlegel2015pi, grinnell2006nested, miron2010differential, kim2013cooperative, moon2021signal}. The inputs could be chemical attractants or repellants, mechanical stimuli, pH, etc; the output could be gene expression, motility, directional migration, etc. A synergistic response is one in which the response to both signals is larger than to either signal individually (Fig. \ref{antagonism}A). Conversely, an antagonistic response is one in which the response to both signals is smaller than to either individually (Fig.\ \ref{antagonism}B). Whereas most cell responses are synergistic \cite{badache2001interleukin, zhou2007synergistic, mosadegh2008epidermal, uttamsingh2008synergistic, buonato2015egf, schlegel2015pi}, some are notably antagonistic \cite{grinnell2006nested, miron2010differential, kim2013cooperative, moon2021signal}, such as the fact several cancer cell types respond more weakly to a combination  of epidermal growth factor and transforming growth factor $\beta$ gradients than to either gradient alone \cite{moon2021signal}. We focus on the antagonistic response because it is counterintuitive and therefore more difficult, in principle, to understand a priori how it is encoded within a signaling network.

We consider three ubiquitous mechanisms by which network components interact: up- or down-regulation, conversion of one species to another, and reversible binding (Fig.\ \ref{method}). In each case, we uncover the minimal network or networks that explain one of two types of antagonism: {``value antagonism''} (antagonism in the output's value, as in gene expression or motility) or {``slope antagonism''} (antagonism in the output's slope, as in directional migration, where the cell responds to signal changes in space or time). For each interaction class and antagonism type, a network structure emerges that is mechanistically interpretable, exhibiting features such as signal saturation, {mutual inhibition}, or sequestration of a molecular species. Moreover, we find commonalities among interaction classes for a given antagonism type: value antagonism generally requires {mutual inhibition} of network pathways, while slope antagonism generally requires the convergence of multiple pathways on a shared network component. We discuss generalizations of our method and apply it to our recently published data on antagonistic response in cancer cell migration \cite{moon2021signal} and to {sugar utilization data in \textit{Escherichia coli} bacteria}.

\section*{METHODS}

\begin{figure*}
\centering
\includegraphics[width=0.8\linewidth]{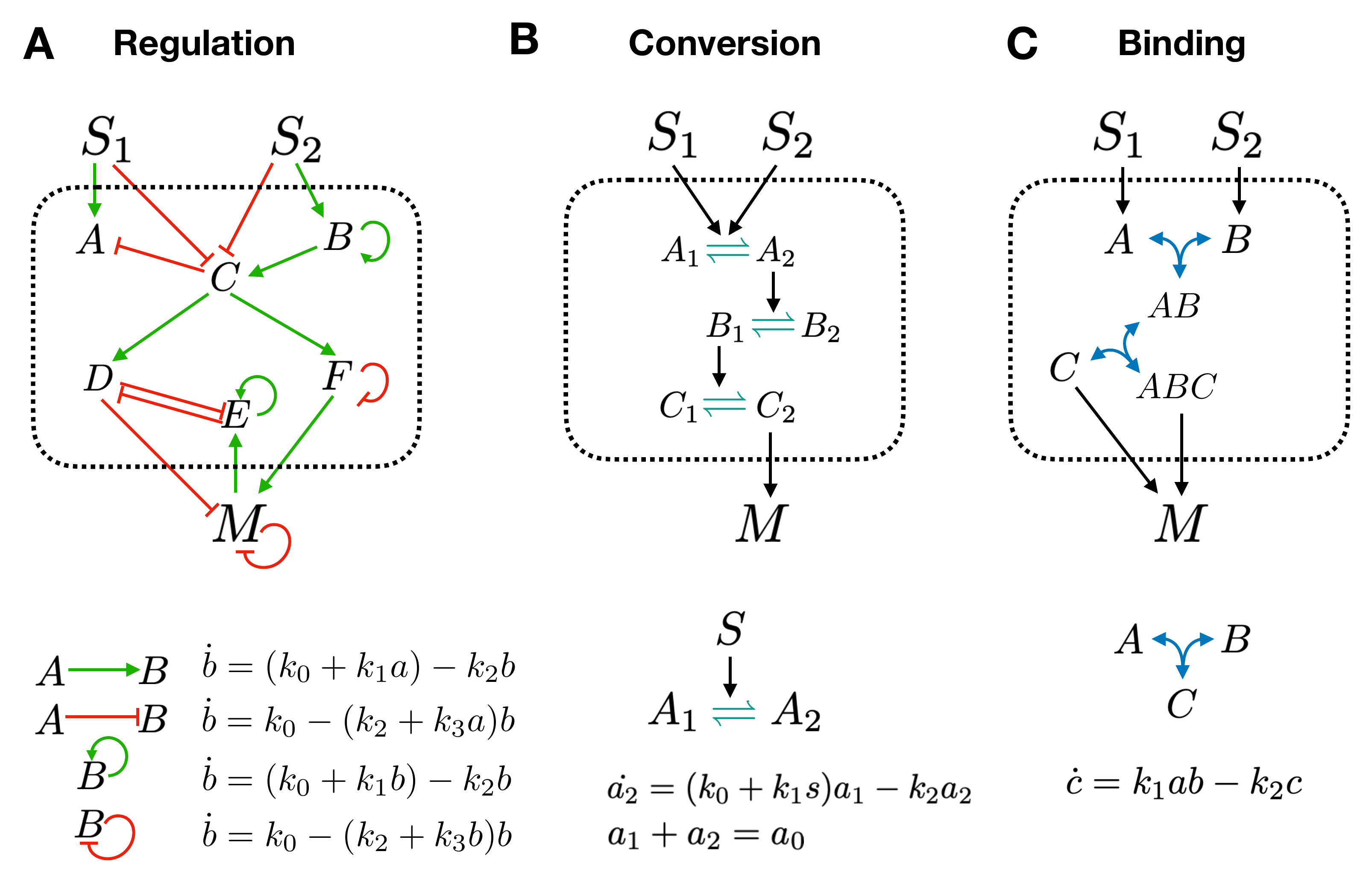}
\caption{\textbf{Mathematical modeling.} For networks containing one of three possible types of interaction, (A) regulation, (B) molecular conversion, or (C) reversible binding, we model the dynamics using deterministic rate equations, as shown. For each network we calculate the steady state $m(s_1,s_2)$, which determines how the output $M$ changes with the inputs $S_1,S_2$. To focus on network topology, we set rates (the $k_i$) and conserved quantities (like $a_0$) to 1 when assessing whether value antagonism or slope antagonism is achieved.}
\label{classes}
\end{figure*}

We consider three interaction types: regulation, conversion, and binding (Fig.\ \ref{classes}). For a given interaction type, we construct networks with two input signals, $S_1$ and $S_2$, and one output species, $M$. The dependence of the output on the inputs, $m(s_1,s_2)$ is given by the steady state of the deterministic rate equations describing the network. For regulation networks (Fig.\ \ref{classes}A), each species undergoes zeroth-order production and first-order degradation. We model up- and down-regulation by making the production rate and degradation rate, respectively, depend linearly on the regulator species (effectively increasing the order of these reactions to first and second, respectively). For conversion networks (Fig.\ \ref{classes}B), a species catalyzes the reversible conversion of a second species into a modified form via a second-order reaction. The total amount of the second species is conserved. For binding networks (Fig.\ \ref{classes}C), two species reversibly combine into a third via a second-order reaction.

{Our choice of linear or bilinear reactions in Fig.\ \ref{classes} is made for simplicity and analytic tractability. Moreover, we will see that it facilitates a focus on network topology, which enables mapping to real biological networks and identification of common features among the interaction classes. Nevertheless, this choice excludes a broad range of nonlinear effects that are relevant in regulatory networks. For example, we will see that mutual inhibition emerges as a key determinant of value antagonism, and only when mutual inhibition is coupled with nonlinear interactions does it enable bistability and therefore mutually exclusive states. Such systems are beyond the scope of the present work.}

{To focus on network topology, we set all parameters ($k_i$ and $a_0$ in Fig.\ \ref{classes}) to $1$ when computing the input-output function $m(s_1,s_2)$.} We consider networks in which the presence of each stimulus individually results in an increased output: $m(s_1,0) > m(0,0)$ and $m(0,s_2) > m(0,0)$.
A network then exhibits value antagonism if there exist value(s) of $s_1$ and $s_2$ for which $m(s_1,s_2) < m(s_1,0)$ and $m(s_1,s_2) < m(0,s_2)$. A network exhibits slope antagonism if there exist value(s) of $s_1$ and $s_2$ for which the slope is positive along each stimulus direction, $\partial m(s_1,0)/\partial s_1>0$ and $\partial m(0,s_2)/\partial s_2>0$, but the slope along the diagonal, $\partial m(s_1,s_2)/\partial s_1+\partial m(s_1,s_2)/\partial s_2$, is less than each of these slopes.

For all mechanisms discussed above, the minimal network(s) can be found exhaustively. All calculations for the network dynamics are performed in Mathematica (see Data Availability).

\section*{RESULTS}

{We first focus on slope antagonism because it will turn out that slope antagonism can be achieved by simpler networks than value antagonism.} Slope antagonism occurs if the derivative of the output with respect to each input, in the absence of the other input, is larger than the derivative when both inputs are present. The derivative of the output with respect to the input is a common response measure for sensing and migration, which require a cell to take temporal or spatial derivatives of a sensory input to determine direction \cite{sourjik2012responding, mugler2016limits, varennes2016sense}.

\begin{figure*}
\centering
\includegraphics[width=.8\linewidth]{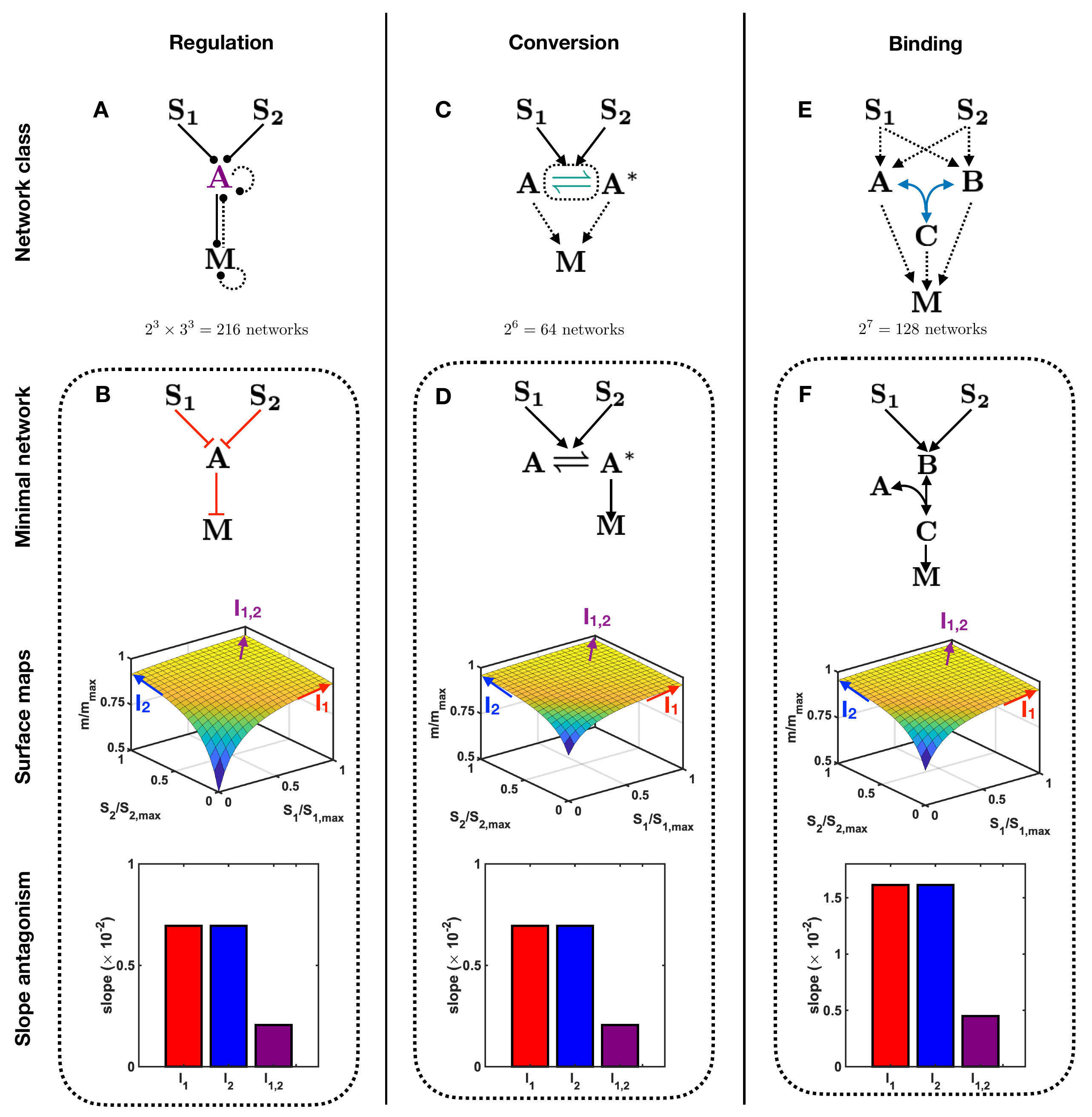}
\caption{\textbf{Slope antagonism.} Slope antagonism occurs if the derivative of the output with respect to each input, in the absence of the other input, is larger than the derivative when both inputs are present. The minimal (A) regulation network class contains (B) a minimal network that achieves slope antagonism by the repression of a repressor. The minimal (C) conversion network class contains (D) a minimal network that achieves slope antagonism by pathway saturation. The minimal (E) binding network class contains (F) a minimal network that achieves slope antagonism by complex formation. In all cases, the output saturates as a function of the inputs (surface maps) and thereby shows slope antagonism (bar graphs). {The values of $I_1$, $I_2$, and $I_{1,2}$ are calculated at $S_i=S_{i,\max}$.}}
\label{slope}
\end{figure*}

\subsection*{In networks with regulation, slope antagonism is minimally achieved by repression of a repressor}

The simplest regulation network with two inputs and one output is a network in which each input either up- or down-regulates the output directly (and the output potentially regulates itself). We find that no network in this class can achieve antagonism (Fig.\ \ref{reg3node}). Therefore we add an intermediate node $A$, which is regulated by each input $S_1$ and $S_2$, regulates the output $M$, and is potentially regulated by the output and itself (Fig.\ \ref{slope}A). Accounting for up-, down-, and potentially absent regulation, this class contains $2^3\times3^3 = 216$ candidate networks. Of these, $2^3 = 8$ candidate networks have the minimum of three regulatory edges. Accounting for the symmetry between the input signals $S_1$ and $S_2$ gives 6 unique networks (Fig.\ \ref{reg4node}). Of these, we find that only one can achieve slope antagonism. This network is shown in Fig.\ \ref{slope}B (top), and its steady-state input-output function (with all parameters set to one) reads
\begin{equation}
\label{reg_slope}
m(s_1,s_2) = \frac{s_1+s_2+1}{s_1+s_2+2}
\end{equation}
(see Data Availability). We see in Fig.\ \ref{slope}B (top) that each input represses the intermediate species, which represses the output. As a result of this double repression, Eq.\ \ref{reg_slope} is an increasing function of $s_1$ or $s_2$ (Fig.\ \ref{slope}B, middle). {However, the repression of intermediate $A$ increases with the increasing levels of inputs ($S_1,S_2$). Owing to this effect, at large levels of inputs, the intermediate molecule $A$ goes to zero, leaving the output $M$ independent of the inputs and saturating to a constant maximum value. This effect is stronger when both inputs are present compared to when just one is present.} Together, this means that the output saturates as a function of the inputs (Fig.\ \ref{slope}B, middle). Thus, the slope is lowest when both inputs are present, i.e.\ the network exhibits slope antagonism (Fig.\ \ref{slope}B, bottom).
{We note that the absolute values in Fig.\ \ref{slope}B (middle and bottom) are an outcome of setting all parameters to one and that in general, the comparison between output values is meaningful in our approach rather than the absolute output values.}

\subsection*{In networks with conversion, slope antagonism is minimally achieved by pathway saturation}

The simplest conversion network with two inputs and one output contains one intermediate species that can be converted between two states, $A$ and $A^*$ (Fig.\ \ref{slope}C). Each input catalyzes one or both conversion directions, and one or both states of the convertible species can regulate the output. Accounting for these possibilities, this class contains $2^6 = 64$ candidate networks. Of these, $2^3 = 8$ candidate networks have the minimum of three regulatory edges. Accounting for symmetry gives three unique networks (Fig.\ \ref{con5node}). Of these three, we find that only one can achieve slope antagonism. This network is shown in Fig.\ \ref{slope}D (top), and its steady-state input-output function (with all parameters set to one) reads
\begin{equation}
\label{con_slope}
m(s_1,s_2) = \frac{2s_1+2s_2+3}{s_1+s_2+2}
\end{equation}
(see Data Availability). We see in Fig.\ \ref{slope}D (top) that both inputs catalyze the conversion of $A$ to $A^*$, and $A^*$ activates the output. Due to conservation of the intermediate species, the presence of both inputs can bring the conversion reaction closer to its saturation point (with all molecules in the $A^*$ state) than either input alone. Thus, the output saturates as a function of the inputs, and the slope is lowest when both inputs are present (Fig.\ \ref{slope}D, middle), i.e.\ the network exhibits slope antagonism (Fig.\ \ref{slope}D, bottom). Although the network and mechanism are different, the effect is the same, and indeed the mathematical expression has the same form, as for the previous network (Fig.\ \ref{slope}B).

\subsection*{In networks with binding, slope antagonism is minimally achieved by complex formation}

The simplest binding network with two inputs and one output contains one binding reaction, in which two intermediate species $A$ and $B$ bind reversibly to form a third species $C$ (Fig.\ \ref{slope}E). Each input activates one or both of $A$ and $B$, and at least one of $A$, $B$, and $C$ activates the output. Accounting for these possibilities, this class contains $2^7 = 128$ candidate networks (Fig.\ \ref{bind6node}). Of these, $2^2\times3 = 12$ candidate networks have the minimum of three regulatory edges, and we find that only one unique network can achieve slope antagonism. This network is shown in Fig.\ \ref{slope}F (top), and its steady-state input-output function (with all parameters set to one) reads
\begin{equation}
\label{bind_slope}
m(s_1,s_2) = \frac{1}{2}\left[s_1+s_2+5-\sqrt{(s_1+s_2+1)^2+4}\right]
\end{equation}
(see Data Availability). We see in Fig.\ \ref{slope}F (top) that both inputs activate $B$, which then binds with $A$ to form a complex $C$, and $C$ activates the output. 
{High levels of inputs lead to a large amount of $B$. Because the level of intermediate $C$ and hence output $M$ is dependent on both $A$ and $B$, in this limit $A$ becomes the limiting factor. Because $A$ is present at a constant level, $M$ thus saturates.} (Fig.\ \ref{slope}F, middle). {This saturating effect is stronger when both inputs are present and hence} the network exhibits slope antagonism (Fig.\ \ref{slope}F, bottom). Here both the mechanism and mathematical expression are different than for the previous two networks (Fig.\ \ref{slope}B and D). Nevertheless, all three networks exhibit saturating input-output functions.

\subsection*{In networks with regulation, value antagonism is minimally achieved by {mutual inhibition}}

We now turn to value antagonism, focusing first on regulation networks. Value antagonism occurs if the value of the output is smaller with both inputs present than with either input alone. Value antagonism is pertinent to non-directional responses, such as molecular abundance, fluorescence level, and other scalar cell properties including speed or size.

\begin{figure*}
\centering
\includegraphics[width=0.8\linewidth]{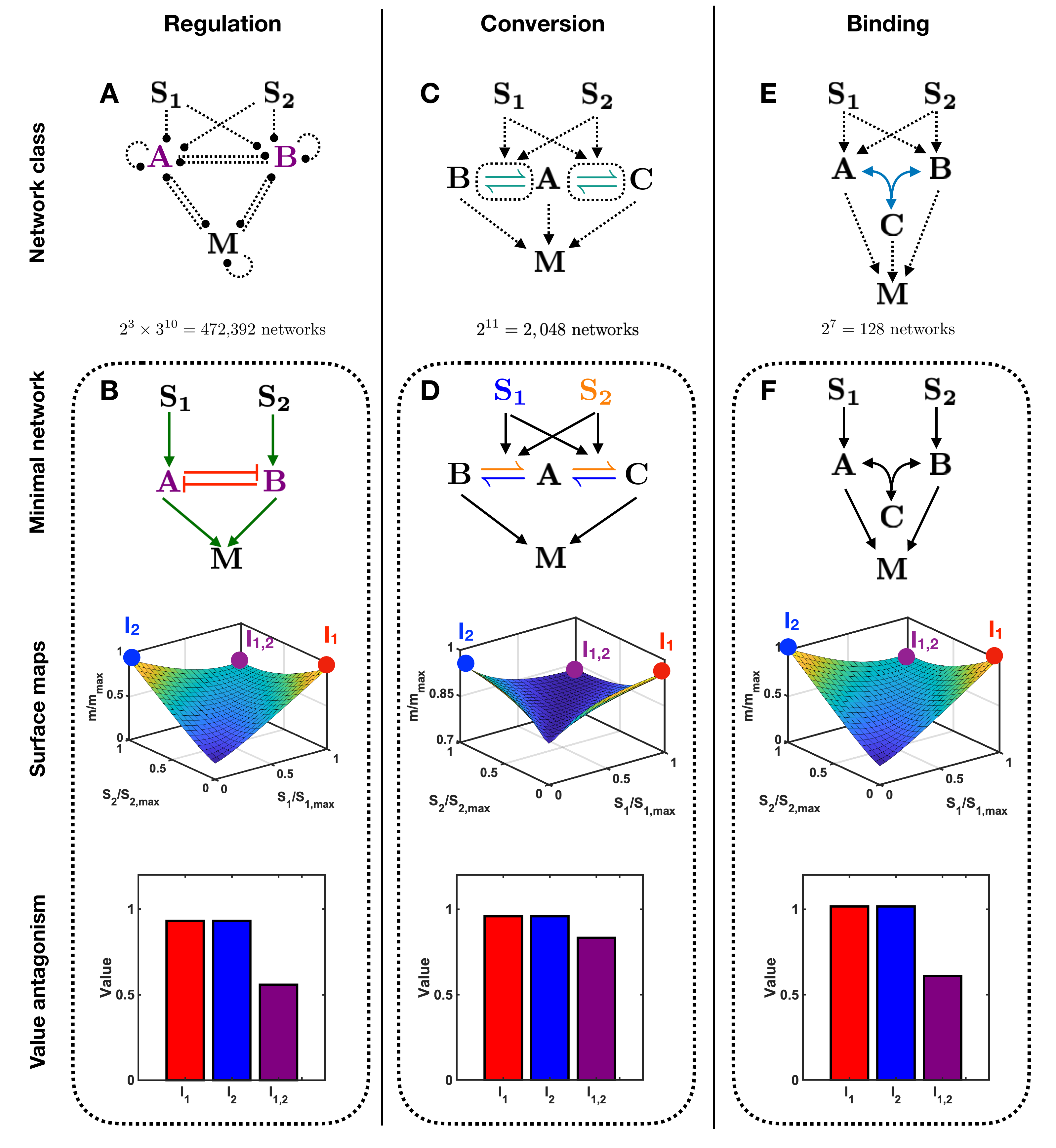}
\caption{\textbf{Value antagonism.} Value antagonism occurs if the value of the output with either input present alone is larger than the value when both inputs are present. The minimal (A) regulation network class contains (B) a minimal network that achieves value antagonism by {mutual inhibition}. The minimal (C) conversion network class contains (D) a minimal network that achieves value antagonism by competing fluxes. The minimal (E) binding network class contains (F) a minimal network that achieves value antagonism by sequestration. In all cases, the output is lower with both inputs present than with either input alone (surface maps) and thereby shows value antagonism (bar graphs). {The values of $I_1$, $I_2$, and $I_{1,2}$ are calculated at $S_i=S_{i,\max}$.}}
\label{value}
\end{figure*}

We find that no network in the class of regulation networks with one intermediate node (Fig.\ \ref{slope}A) exhibits value antagonism. Therefore we add a second intermediate node $B$ (Fig.\ \ref{value}A). Accounting for the additional ways that $A$ and $B$ can regulate the output and each other, this class contains $2^3\times3^{10} = 472{,}392$ candidate networks.  We systematically analyze five node regulatory networks with three edges (2,288 possible networks), four edges (11,440 possible networks), five edges (41,184 possible networks) and six edges (109,824 possible networks; see Data Availability). None of the possible networks with three, four, or five edges show value antagonism. Accounting for symmetry in networks, our analysis gives five unique networks with six regulatory edges that satisfy the value antagonism condition (Fig.\ \ref{reg5node}). Three of the five networks (Fig.\ \ref{reg5node}A-C), a representative one of which is shown in Fig.\ \ref{value}B (top), exhibit {mutual inhibition} (the other two networks are less intutitive to interpret but can be analyzed in detail in future studies). In {mutual inhibition}, each input signal activates an intermediate node and, either directly or via the intermediate node, represses the other intermediate node. In Fig.\ \ref{value}B (top), one input activates $A$ while the other activates $B$, and $A$ and $B$ repress each other. Its steady-state input-output function (with all parameters set to one) reads
\begin{equation}
\label{reg_value}
m(s_1,s_2) = \sqrt{s_1^2-2 \left(s_2-1\right) s_1+\left(s_2+1\right){}^2+4}
\end{equation}
(see Data Availability). When either input is present alone, it produces high output via its activated intermediate. However, when both inputs are present together, the {mutual inhibition} results in neither intermediate being strongly produced, and thus a low output. Consequently, the value of the output is lower with both inputs present than with either alone (Fig.\ \ref{value}B, middle), i.e.\ the network exhibits value antagonism (Fig.\ \ref{value}B, bottom). {We reiterate that our conclusions are restricted to the choice of linear interactions and exclude nonlinear effects such as bistability that often coincide with mutual inhibition such as that in Fig.\ \ref{value}B.}

\subsection*{In networks with conversion, value antagonism is minimally achieved by competing fluxes}

We find that no network in the class of conversion networks with one intermediate convertible species (Fig.\ \ref{slope}C) exhibits value antagonism. Therefore we add a second reversible conversion reaction. The simplest way to do so (i.e.\ with the fewest added nodes) is to allow the intermediate species to be converted among three states, $A$, $B$, and $C$ (Fig.\ \ref{value}C). Allowing each input to catalyze either conversion in either direction, and ensuring that the output is regulated by at least one intermediate state, this class contains $2^{11} = 2{,}048$ candidate networks. Of these, we find that only one unique network with the minimum six regulatory edges can achieve value antagonism (Fig.\ \ref{con6node}). This network is shown in Fig.\ \ref{value}D (top), and its steady-state input-output function (with all parameters set to one) reads
\begin{equation}
\label{con_value}
m(s_1,s_2) = \frac{2s_1^2+2s_2^2+s_1s_2+5s_1+5s_2+5}{s_1^2+s_2^2+s_1s_2+3s_1+3s_2+3}
\end{equation}
(see Data Availability). We see in Fig.\ \ref{value}D (top) that one input catalyzes both conversion reactions in one direction, while the other input catalyzes both conversion reactions in the other direction. When either input is present alone, the flux in state space is driven to one extreme ($B$ or $C$), and that state activates the output. However, when both inputs are present together, the fluxes compete, and the central state $A$ attains appreciable occupancy. Because only the extreme states activate the output, the output production goes down. Consequently, the value of the output is lower with both inputs present than with either alone (Fig.\ \ref{value}D, middle), i.e.\ the network exhibits value antagonism (Fig.\ \ref{value}D, bottom).

\subsection*{In networks with binding, value antagonism is minimally achieved by sequestration}

We find that in the class of binding networks with one binding reaction (Fig.\ \ref{slope}E, also shown in Fig.\ \ref{value}E), no network with the minimum three edges exhibits value antagonism. Therefore we consider networks in the class with four edges. Of the $7{\rm -choose-}4 = 7!/(3!4!) = 35$ four-edge networks, we find that only one unique network can achieve value antagonism (Fig.\ \ref{bind6node}). This network is shown in Fig.\ \ref{value}F (top), and its steady-state input-output function (with all parameters set to one) reads
\begin{equation}
\label{con_value}
m(s_1,s_2) = \sqrt{s_1^2+s_2^2-s_1s_2+2s_1+2s_2+5}
\end{equation}
(see Data Availability). We see in Fig.\ \ref{value}F (top) that one input activates the output through $A$, and the other input activates the output through $B$. If only one input is present, only  $A$ or $B$, but not both, is produced, and minimal binding occurs. However, if both inputs are present, both $A$ and $B$ are produced, and  strong binding occurs. Strong binding removes molecules from the $A$ and $B$ states and therefore reduces the production of $M$. Thus, the bound $C$ state sequesters molecules from the $A$ and $B$ states, reducing the production of $M$. Sequestration results in the output taking a smaller value in the presence of both inputs than in the presence of either alone (Fig.\ \ref{value}F, middle), and the network exhibits value antagonism (Fig.\ \ref{value}F, bottom).

\subsection*{Common structural features among minimal networks}

The minimal networks that exhibit slope antagonism (Fig.\ \ref{slope}B, D, and F, top) share a common structure, despite the different interaction types from which they are built. Specifically, each involves both inputs converging on a common pathway (a regulatory species, a conversion reaction, or a bound complex, respectively), which then regulates the output. We conjecture that this convergent pathway structure is a general minimal requirement for slope antagonism because two inputs converging onto a single pathway can lead to saturation of that pathway when both inputs are present. As a result, we predict that cell responses that show slope antagonism will be governed by a signaling network that exhibits some degree of pathway convergence.

Similarly, the minimal networks that exhibit value antagonism (Fig.\ \ref{value}B, D, and F, top) share a second common structure. Specifically, each involves two output-activating pathways, each one at cross-purposes with the other. In the regulation network (Fig.\ \ref{value}B, top), each pathway down-regulates the other via intermediate nodes. In the conversion network (Fig.\ \ref{value}D, top), each pathway provides a competing flux against the other. In the binding network (Fig.\ \ref{value}F, top), each pathway sequesters molecules from the other. We conjecture that this {mutual inhibition} structure is a general minimal requirement for value antagonism because two inputs at cross-purposes can lead to a reduced output when both inputs are present. As a result, we predict that cell responses that show value antagonism will be governed by a signaling network that exhibits some degree of {mutual inhibition}.

\begin{figure*}
\centering
\includegraphics[width=0.8\linewidth]{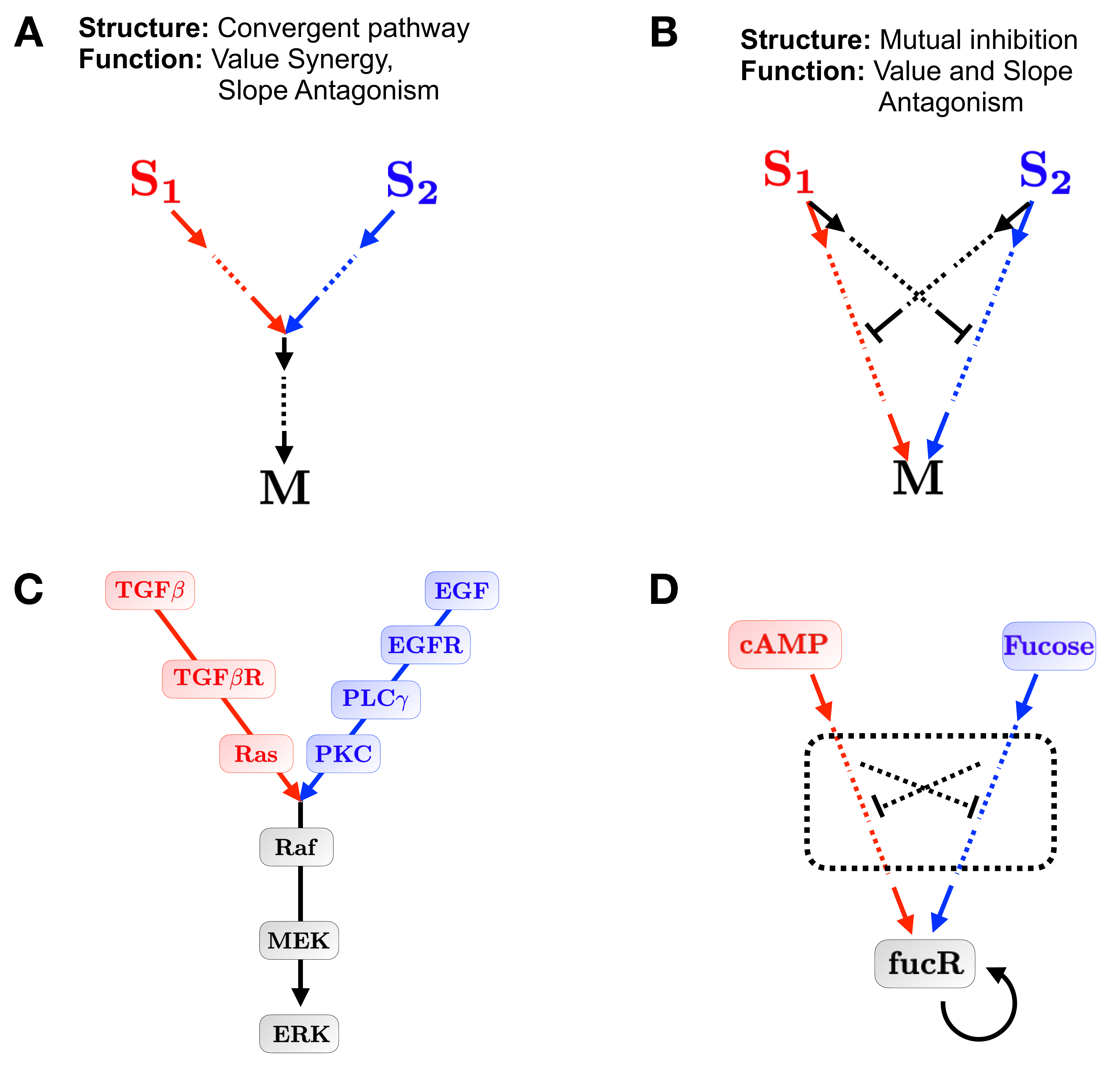}
\caption{\textbf{Common structural features.} (A) The minimal networks that exhibit slope antagonism (Fig.\ \ref{slope}B, D, and F, top) also exhibit value synergy and have a common structure of two converging pathways. (B) The minimal networks that exhibit value antagonism (Fig.\ \ref{value}B, D, and F, top) also exhibit slope antagonism and have a common structure of two mutually {inhibitory} pathways. (C) The network that integrates EGF and TGF-$\beta$ in cancer cells shows pathway convergence, consistent with migration experiments on cancer cell lines indicating value synergy and slope antagonism \cite{moon2021signal}. Network adapted from \cite{wang2009cross}.
(D) {The fucR gene exhibits value antagonism in response to fucose and cAMP \cite{kaplan2008diverse}. Our analysis predicts that mutually inhibitory interactions should exist in the network downstream of fucose and cAMP and upstream of fucR.}}
\label{common}
\end{figure*}

Finally, we have found that the minimal networks that exhibit slope antagonism do so without exhibiting value antagonism; instead they exhibit value synergy (Fig.\ \ref{slope}B, D, and F, middle). In contrast, the minimal networks that exhibit value antagonism also exhibit slope antagonism (Fig.\ \ref{value}B, D, and F, middle). Therefore, taken together, our results suggest that cell responses that show value synergy but slope antagonism will likely be governed by a signaling network with convergent pathways (Fig.\ \ref{common}A), whereas cell responses that show value antagonism and slope antagonism will likely be governed by a signaling network with mutually {inhibitory} pathways (Fig.\ \ref{common}B).

\subsection*{Applications to cancer cell migration and bacterial metabolism}

Finally, we demonstrate our method by applying it to two case studies. {Our applications are presented as proofs of principle. Comparison of our minimal networks to biological networks is valid under the assumption that either (i) the minimal network represents a coarse-graining of the larger biological network, in which each of our nodes represents multiple signaling components; or (ii) the biological network is modular, such that the behavior of a subnetwork can be understood separately from the rest of the network \cite{hartwell1999molecular, alon2006introduction}.}

First, we consider our recently published data on the migratory response of cancer cell lines to two growth factor gradients \cite{moon2021signal}. We found, for both a breast cancer and pancreatic cancer cell line, that cells migrated in the direction of an epidermal growth factor (EGF) gradient and a transforming growth factor $\beta$ (TGF-$\beta$) gradient when each gradient was presented individually. However, cells' directional migration accuracy was weaker when both gradients were presented together. Because directional migration requires detecting a derivative, this set of antagonistic responses corresponds specifically to slope antagonism. Furthermore, we found that each factor increased the average speed of cells individually, and that the speed did not decrease when both factors were presented together. Because speed is a scalar measure, this set of responses rules out value antagonism. Our analysis here suggests that the presence of two converging pathways as shown in Fig.\ \ref{common}A is necessary to explain the experimental results.

Indeed, using the conversion mechanism for slope antagonism (Fig.\ \ref{slope}D) in our previous work \cite{moon2021signal}, we successfully predicted the experimental response of the cells to different input conditions, including combinations of graded and uniform signals. In that work, the mathematical model was introduced {\it ad hoc}, and its success was fortuitous: it was just one possible description among many, in principle. Here, in contrast, we have shown that this model corresponds to the class of unique minimal signaling network that can explain the experimental data.

Furthermore, turning to the known pathways by which EGF and TGF-$\beta$ are processed in cancer cells, we find that the pathway structure resembles that of Fig.\ \ref{common}A. Specifically, as seen in Fig.\ \ref{common}C (adapted from \cite{wang2009cross}), the EGF receptor activates PLC$\gamma$ and PKC, while the TGF-$\beta$ receptor activates Ras; both of these pathways then converge on the Raf-MEK-ERK cascade. While this particular network has been identified in the context of cell cycle control in non-small cell lung cancer \cite{wang2009cross}, the Raf-MEK-ERK cascade has been shown to play a central role in cancer cell migration as well \cite{katagiri2010mek, ehrenreiter2005raf}. {Although the Raf-MEK-ERK cascade is known to act as a convergence point for more than just these two pathways (e.g., it is also activated by FGF receptors \cite{yamazaki2010both}), other stimuli are absent in the above experiments, suggesting that these other pathways would not be activated. We expand on multi-input networks in the Discussion.}

Second, we consider {known regulatory networks that exhibit value antagonism. Although such networks are argued to be possible in principle \cite{buchler2003schemes, hermsen2006transcriptional} and are commonly constructed synthetically \cite{tamsir2011robust, moon2012genetic}, natural examples are notably rare. An exception is the fucR gene network in \textit{E.\ coli} bacteria. The fucR gene is activated individually by the sugar fucose and the small molecule cyclic adenosine monophosphate (cAMP), but when both inputs are present together the fucR response decreases \cite{kaplan2008diverse}. The fucR network thus exhibits value antagonism. Our analysis here suggests that the presence of mutual inhibition as shown in Fig.\ \ref{common}B is expected to explain this response.}

{However, apart from the activation by fucose and cAMP, we are aware of only one regulatory interaction involving fucR: it self-activates \cite{tierrafria2022regulondb, keseler2021ecocyc}. Thus, we predict the presence of hitherto unknown mutually inhibitory interactions downstream of fucose and cAMP and upstream of fucR (Fig.\ \ref{common}D). Most biological networks are incompletely characterized. Indeed, even in the {\it E.\ coli} gene regulatory network the regulation of approximately 65\% of promoters remains unknown \cite{ireland2020deciphering}. Therefore, it is plausible that regulatory interactions upstream of fucR are missing. Our results provide a specific guide for the nature of these interactions.}

These two examples illustrate that our formalism not only predicts the minimal network that can explain complex cell behavior, it also facilitates identification of key components from pathway data and predicts minimal structures to seek {when not all interactions are known.}

\section*{DISCUSSION}

Biochemical networks are inherently complex. This complexity prevents intuitive understanding and makes it difficult to generate unique, falsifiable predictions for experiments. Here we have turned this problem around. We have developed a method to {deduce} the minimal biochemical network consistent with experimental observations. We have demonstrated our method on the counterintuitive observation of antagonism, where the response of a cell to two signals is weaker than the response to either signal alone. Our method has revealed six intuitive mechanisms, corresponding to networks of three interaction types (regulation, conversion, and binding), to explain two types of antagonism (in the value and in the slope of the output).
{For each antagonism type, we find that the minimal networks are structurally similar. Specifically, networks exhibiting slope antagonism contain two convergent pathways. Networks exhibiting value antagonism contain two mutually inhibitory pathways.} Applying our method to two examples in which slope or value antagonism is observed experimentally, we conclude that a convergent pathway structure or {mutual inhibition} should be present in the underlying network, respectively. {In one case this conclusion is} indeed consistent with the known biochemical data, {and in the other it serves as a prediction for a subnetwork that is sparsely characterized.}

The number of signaling molecules or pathway component molecules are integer-valued variables. We have approximated their concentrations as real-valued variables when testing for synergy or antagonism in their input-output functions. Nevertheless, synergy and antagonism in continuous functions have analogs in the reduced representation of binary logic gates. Specifically, synergy corresponds to monotonic logic gates such as OR or AND, where the output with both inputs present is greater than or equal to, respectively, that with either input present alone. Antagonism corresponds to XOR, where the output is low with both inputs present or absent, but high with either input present alone. Binary logic gates have been demonstrated in cell networks, including gene regulatory networks \cite{setty2003detailed, buchler2003schemes, kaplan2008diverse} or even single molecules \cite{de2012protein}, and indeed gate formation and concatenation form the basis of much of synthetic biology \cite{tamsir2011robust, moon2012genetic, shis2014modular}.

{We perform all our analysis at the mean level of the output, ignoring noise. Noise is ubiquitous in biochemical signaling networks \cite{raj2008nature, elowitz2002stochastic}, raising the question of how noise may affect our results. The presence of noise would introduce an additional burden on distinguishing high from low values, or high from low slopes, in our analysis. However, we suspect that accounting for noise would simply imply a stricter condition on the parameter regime associated with value or slope antagonism for a given topology, rather than changing the deduction of the minimal topology itself. On the other hand, the consideration of noise opens the possibility for new objective measures, beyond the value or the slope of the output variable as used here. For example, it has been shown that the precision with which a slope can be estimated decreases with the background concentration if the slope is held fixed \cite{mugler2016limits}. Therefore, this precision may exhibit saturation with two inputs without the network itself inducing saturation in the mean output. We leave the investigation of noise-based output measures to future work.}

We have developed our method in the context of a cell responding to two chemical signals. However, cells respond to a variety of signal types, including chemical signals, temperature, fluid flow, stiffness, and more. Because each of these environmental cues ultimately triggers an internal signaling network in the cell, our formalism of {deducing} minimal networks is equally applicable to any kind of external signal that induces a cellular response. {Furthermore, our method is generalizable to more than two signals. Investigating more than two inputs will lead to a much larger number of possible networks, but it might also give rise to richer and novel functional responses. Alternatively, it might identify common motifs that apply regardless of the number of inputs; for example, pathway convergence ought to lead to saturation and thus slope antagonism regardless of the number of pathways.} Investigating the effect of multiple cues will be important for understanding cells' integrated responses to complex environments {\it in vivo}.

{Our work focuses on amplitude information (input and output levels) and neglects spatial effects. Other studies have investigated the basic networks \cite{ma2006robustness, cotterell2010atlas, schaerli2018synthetic} or decoders \cite{dubuis2013positional, petkova2019optimal, bauer2021trading, shen2021scaling} required to form spatial patterns in embryonic development, particularly in the fruit fly \textit{Drosophila}. In \textit{Drosophila}, a possible connection to value antagonism may be the mutual repression often accompanying stripe formation \cite{small1991transcriptional, small1992regulation, stanojevic1991regulation, jaeger2011gap}. As a concrete example, the Kr\"uppel (Kr) gene is unexpressed in the presence of either the Bicoid (Bcd) morphogen in the anterior or the Caudal (Cad) morphogen in the posterior but expressed in the presence of both morphogens in the middle \cite{jaeger2011gap} (this is the XNOR version of antagonism's XOR logic). Indeed, Bcd and Cad respectively regulate Kr via the intermediate genes Hunchback (Hb) and Knirps (Kni), among others, and Hb and Kni repress each other \cite{jaeger2011gap}, consistent with our predicted picture of value antagonism (Fig.\ \ref{common}B). It is important to point out, however, that this mutual inhibition may serve alternative functions in a spatial context, such as sharpening the spatial boundaries \cite{sokolowski2012mutual}. Equally important is that these interactions are just part of the \textit{Drosophila} gap gene network, which is highly interconnected \cite{jaeger2011gap}, and therefore it may be misleading to isolate these interactions from the whole.}

As illustrated by the presented examples, our method can be used to elucidate the coarse structure of well-characterized networks that is likely responsible for observed cellular behaviors. On the other hand, for less well-characterized networks, our method can generate predictions for these basic structures, given observed cellular behaviors. These predictions can complement biochemical studies, providing clues for what types of interactions may be missing. In this way, our work serves as a top-down guide for more detailed bottom-up network construction.

\section*{ACKNOWLEDGMENTS}
This work was supported by the National Science Foundation (PHY-1945018, MCB-1936761, MCB-2134603), the National Institutes of Health (U01 HL143403, R01 CA254110, U01 CA274304), and the Purdue Center for Cancer Research (P30 CA023168).

\section*{DATA AVAILABILITY}
All calculations for the network dynamics are performed in Mathematica. All code, including pdfs of the Mathematica files, can be found at \url{https://github.com/souticksaha21/Inference-of-signaling-mechanism-from-cellular-responses-to-multiple-cues-Version-2}.

\section*{COMPETING INTERESTS}
The authors declare no competing financial or non-financial interests.

\section*{AUTHOR CONTRIBUTIONS}
SS conceived the idea. SS and AM developed the methods. SS performed the research. All authors analyzed the data. All authors wrote the paper.

\renewcommand{\thefigure}{S\arabic{figure}}
\setcounter{figure}{0}

\begin{figure*}
\centering
\includegraphics[width=0.95\linewidth]{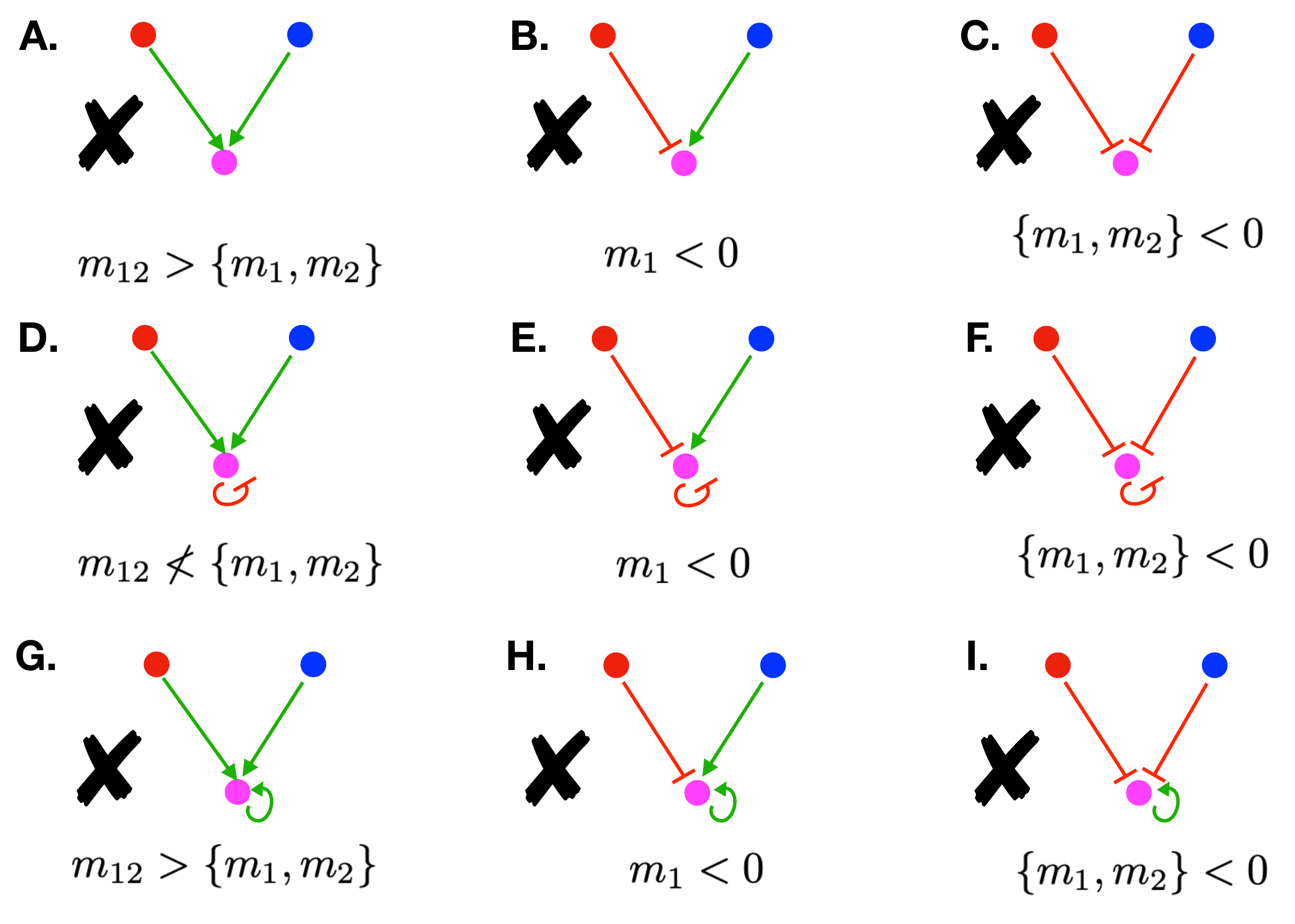}
\caption{\textbf{Slope antagonism with three-node regulatory networks.} Accounting for symmetry there are nine three-node regulatory networks. No three node network can satisfy the slope antagonism condition. The reasons for not satisfying the slope antagonism condition are given under each network. Here, $m_{12}=\partial m(s_1,s_2)/\partial s_1+\partial m(s_1,s_2)/\partial s_2, m_1=\partial m(s_1,0)/\partial s_1, m_2=\partial m(0,s_2)/\partial s_2$.Calculations are performed in Mathematica (see Data Availability).}
\label{reg3node}
\end{figure*}

\begin{figure*}
\centering
\includegraphics[width=0.95\linewidth]{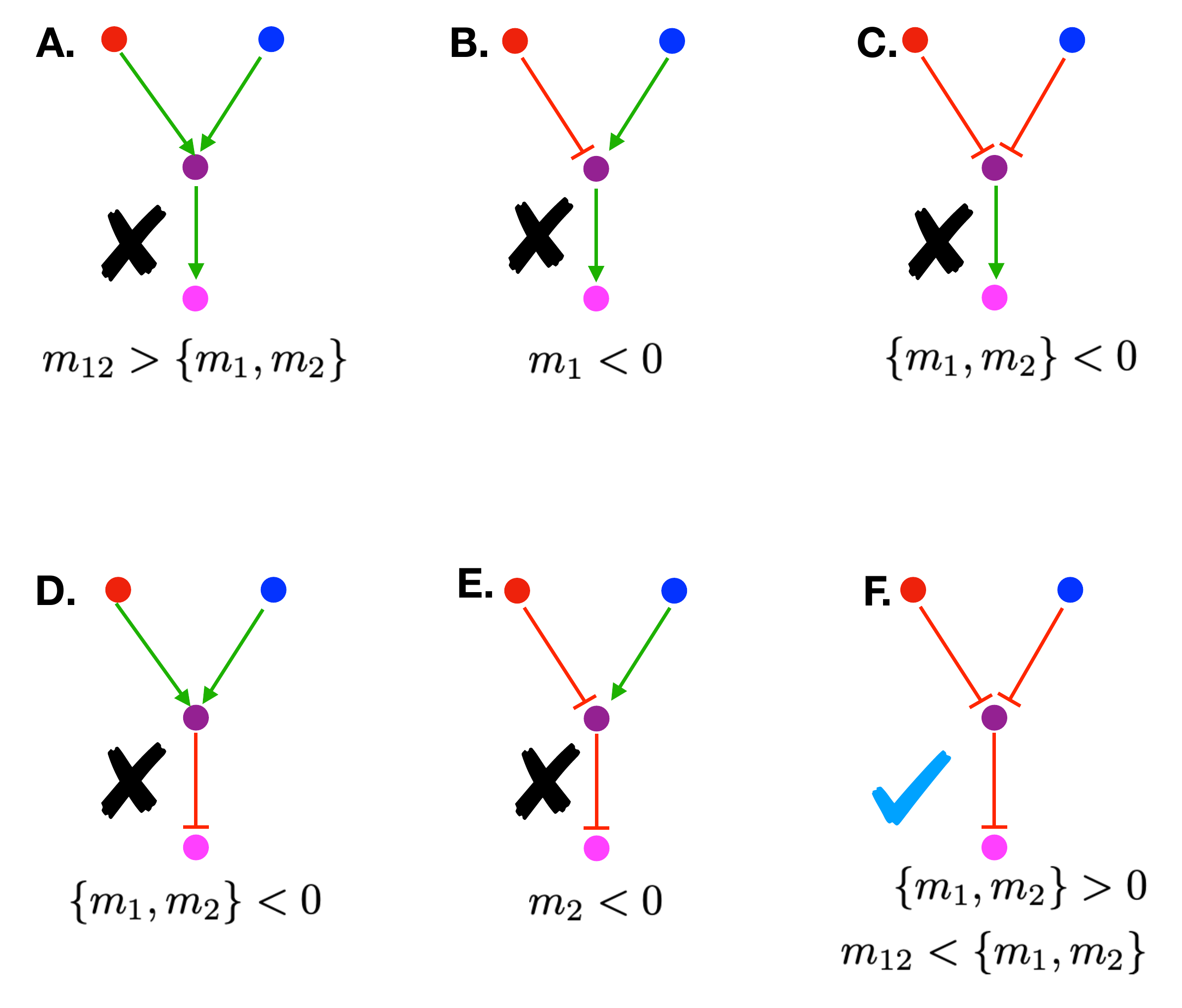}
\caption{\textbf{Slope antagonism with four-node regulatory networks.} Accounting for symmetry there are six four-node regulatory networks with the minimum of three edges. Only one, F, can satisfy the slope antagonism condition. The reasons for satisfying or not satisfying the slope antagonism condition are given under each network. Here, $m_{12}=\partial m(s_1,s_2)/\partial s_1+\partial m(s_1,s_2)/\partial s_2, m_1=\partial m(s_1,0)/\partial s_1, m_2=\partial m(0,s_2)/\partial s_2$. Calculations are performed in Mathematica (see Data Availability).}
\label{reg4node}
\end{figure*}

\begin{figure*}
\centering
\includegraphics[width=0.95\linewidth]{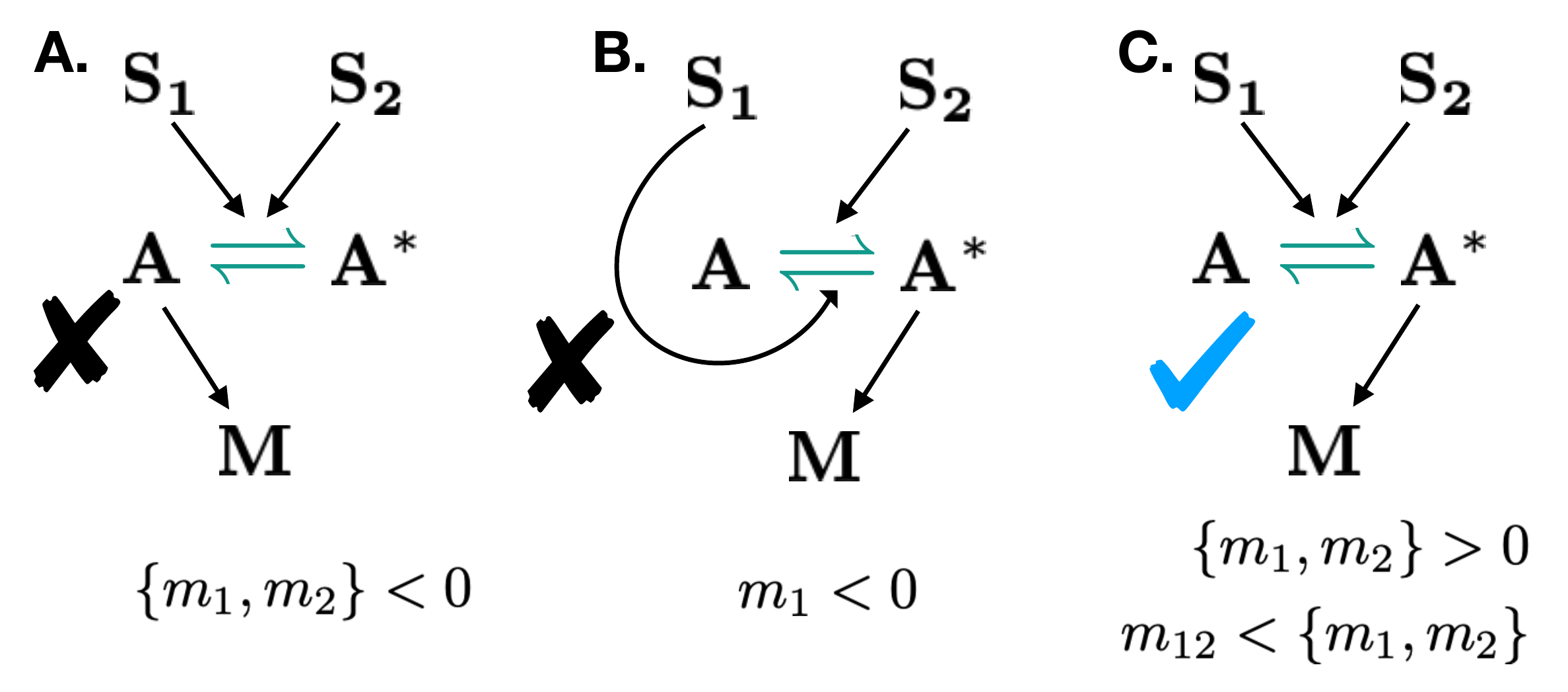}
\caption{\textbf{Slope antagonism with conversion networks.} Accounting for symmetry there are three conversion networks with the minimum of five nodes and three edges. Only one, C, can satisfy the slope antagonism condition. The reasons for satisfying or not satisfying the slope antagonism condition are given under each network. Here, $m_{12}=\partial m(s_1,s_2)/\partial s_1+\partial m(s_1,s_2)/\partial s_2, m_1=\partial m(s_1,0)/\partial s_1, m_2=\partial m(0,s_2)/\partial s_2$. Calculations are performed in Mathematica (see Data Availability).}
\label{con5node}
\end{figure*}

\begin{figure*}
\centering
\includegraphics[width=0.95\linewidth]{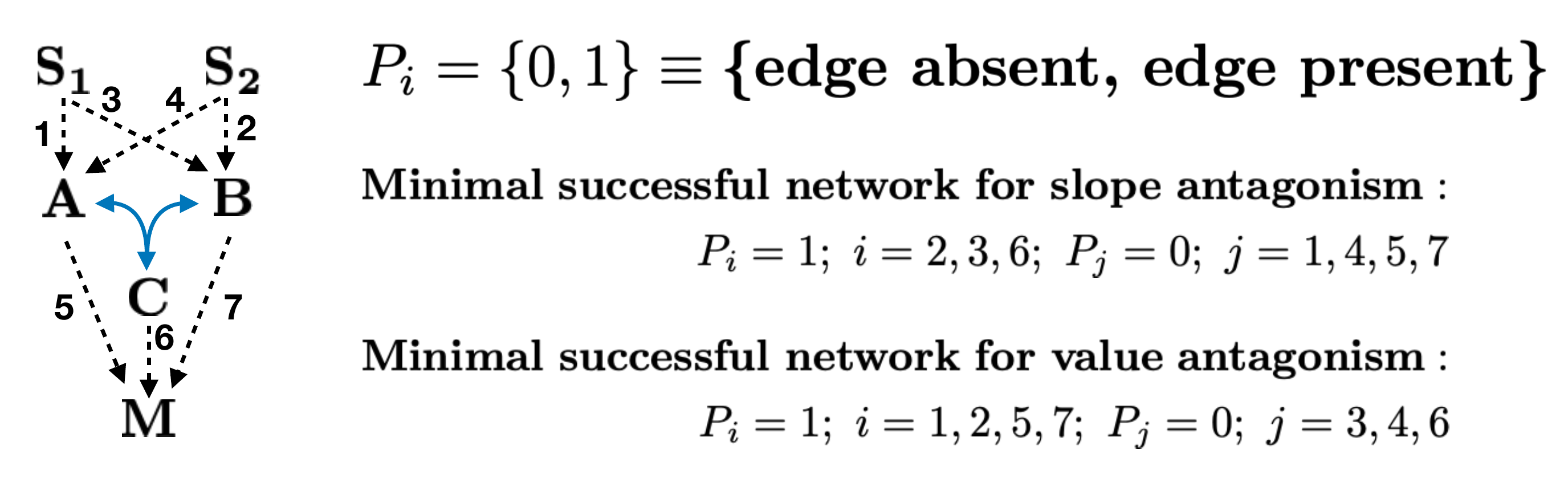}
\caption{\textbf{Six-node binding networks.} There are $2^7 = 128$ six-node binding networks, depending on the presence or absence of the seven edges. The minimal network that can satisfy the slope antagonism condition has three edges (2, 3, and 6) and is shown in Fig.\ \ref{slope}F. The minimal network that can satisfy the value antagonism condition has four edges (1, 2, 5, and 7) and is shown in Fig.\ \ref{value}F. Calculations are performed in Mathematica (see Data Availability).}
\label{bind6node}
\end{figure*}


\begin{figure*}
\centering
\includegraphics[width=0.95\linewidth]{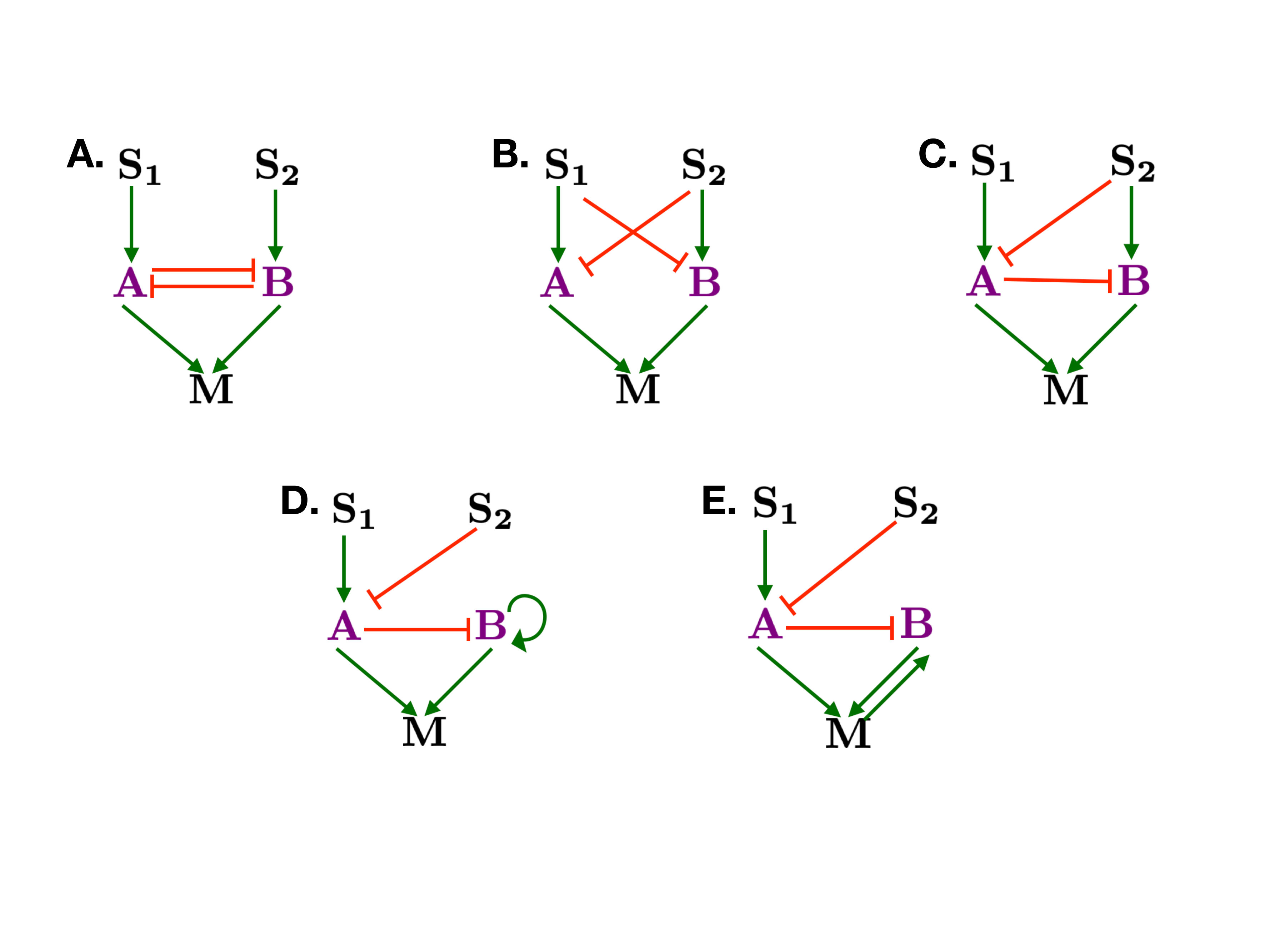}
\caption{{\textbf{Value antagonism with regulation networks.} Five unique five-node, six-edge regulation networks are found using the our mathematical framework to be capable of showing value antagonism. Calculations are performed in Mathematica (see Data Availability).}}
\label{reg5node}
\end{figure*}


\begin{figure*}
\centering
\includegraphics[width=0.95\linewidth]{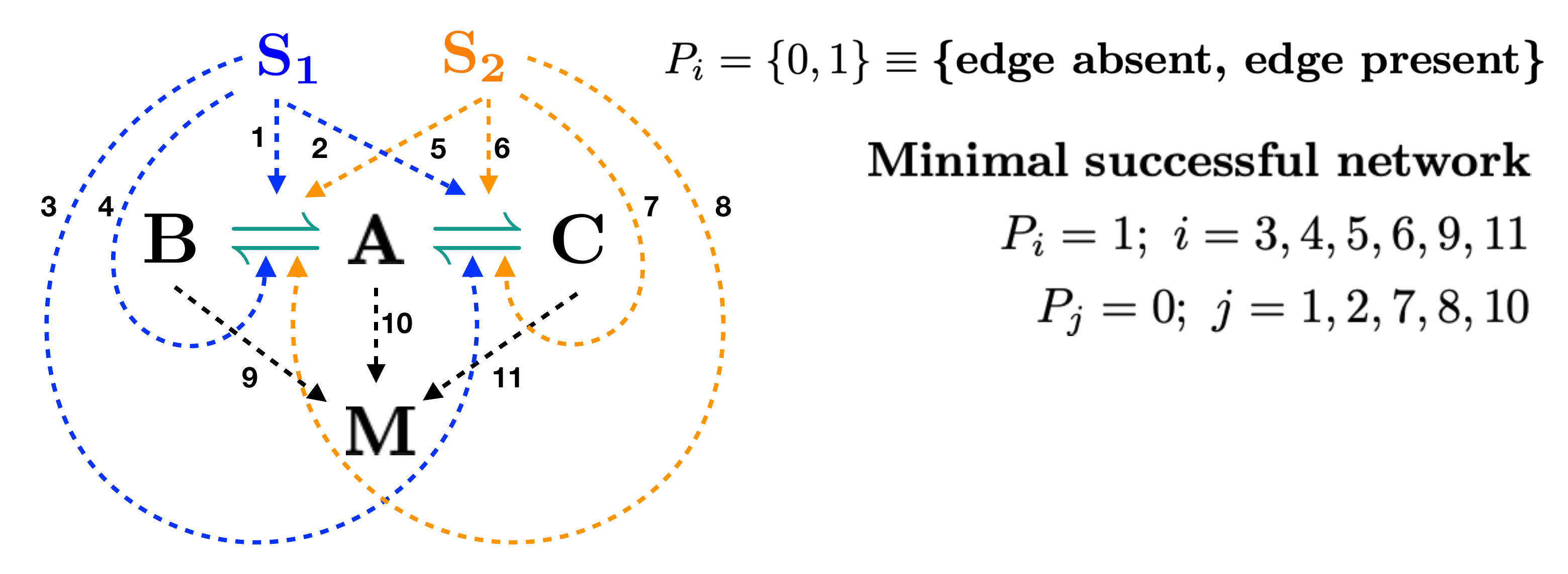}
\caption{\textbf{Value antagonism conversion networks.} There are $2^{11} = 2{,}048$ six-node conversion networks, depending on the presence or absence of the eleven edges. The minimal network that can satisfy the value antagonism condition has six edges (3, 4, 5, 6, 9, and 11) and is shown in Fig.\ \ref{value}D. Calculations are performed in Mathematica (see Data Availability).}
\label{con6node}
\end{figure*}

\end{document}